# LINEAR AND NON-LINEAR MODELS FOR MASTER SCHEDULING OF DYNAMIC RESOURCES PRODUCT MIX

## AYMAN R. MOHAMMED[1,2], AHMAD ABU SLEEM[1,2], MOHAMMAD A. M. ABDEL-AAL[1,2*]

[1]Industrial & Systems Engineering Department
King Fahd University of Petroleum and Minerals, Dhahran 31261, Saudi Arabia
[2]Interdisciplinary Research Center of Smart Mobility & Logistics,
King Fahd University of Petroleum and Minerals, Dhahran 31261, Saudi Arabia
Emails: ahmad_yousef_abu_sleem@yahoo.com; mabdelaal@kfupm.edu.sa;
m.abdelaal82@gmail.com
ORCID: 0000-0001-7068-0089 (A.R. Mohammed), 0000-0002-0919-9699 (M.A.M. Abdel-Aal)

## ABSTRACT

The literature on master production scheduling for product mix problems under the Theory of Constraints (TOC) was considered by many previous studies. Most studies assume a static resources availability. In this study, the raw materials supplied to the manufacturer is considered as dynamic depending on the results of the problem. Thus, an integer linear heuristic, an integer non-linear optimization model, and a basic non-linear model are developed to find a good solution of the problem. The results of the three models were compared to each other in terms of profit, raw materials costs, inventory costs and raw materials utilization. Recent studies in the field are reviewed and conclusions are drawn.

## 1    INTRODUCTION:

In recent years, business decision-makers realized that they need to highly increase their product mix to finely satisfy the targeted customer segment with accelerating the lead times [1]. Thus, the probability of being a good competitor, having sustainable profit, and satisfying the target market increases by producing various products [2]. Non-financial key performance indicators (KPIs) such as customer service, inventory work in process (WIP), and manageability are also affected directly by the product-mix decision [3]. In one way or another, the decision on product mix is extremely important for a firm to be successful [4].

The detailed planning procedure that measures factory production and connects it to orders is defined as master scheduling. The intuition and expertise of the planner in creating realistic master production schedules, selecting acceptable lot sizes for components, and adjusting capacity levels are very important processes [34]. Material and capacity plans should be integrated into numerous industrial environments, taking into account bottleneck places, order penetration points, and material profiles [35]. Manufacturing facilities use master scheduling as a communication tool. This aids in the prioritization of supply chain requirements by examining all requirements and determining what needs to be accomplished first. Having a consistent flow of production is also important. It will provide everyone with a plan of what will be produced and how much capacity is available to make extra things, making it easier to give realistic lead time estimations to clients.

The soft variety product mix problem emerges from the difficulty of assigning the limited supply of resources to various products with satisfying the demand for each product in addition to maximizing the profit [5], [6], [7]. It is a complex combinatorial NP-hard problem, in many cases, it requires the application of artificial intelligence techniques and heuristics to solve it in real-time [8], [9]. It is important to note that the shortage of resources limits the chances of satisfying the demand [10], [11] while unrestrained amounts of resources result in lowering utilization. The resources' utilization in this study is constraint-based as it can't be less than 90%. The raw materials' amounts are flexible depending on the economies of scale and the demand for each product.

Product-mix problems can be solved under the theory of constraints using the 5 steps of focusing developed by Goldratt [12]. The product-cost paradigm of accounting and TOC came up with three interrelated performance measures which are the throughput or the profit, inventory which is the investment in selling products, and expenses of operation [13]. The most important measure of (TOC) is not the number of products to be produced, but the overall throughput or revenue [14]. TOC allows the calculation of the overall marginal profit for the available product-mix solutions under a set of constraints that the decision-maker could select the combination of products that maximize the profit. It is commonly referred to select the product mix with shared costs to allow the maximum economies of scale [15]. This approach is reported to have significant improvement in production performance [16].

Product-Mix master scheduling can be solved using a variety of models, one of which being two-stage stochastic integer programming approaches on moving horizons [36]. Creating a master aggregate scheduling model is another option [37]. In addition to data mining tasks and approaches in production planning and scheduling that focused on short- to mid-range or long-range planning [38]. The effectiveness of these models has been demonstrated in several studies, and we will utilize the integer linear programming mathematical model in our research.

An optimization mathematical model is considered by calculating the optimal output based on exact data on raw materials amounts, their consumption rate, and cost [2]. Integer linear programming (ILP) is one of the methods used for product-mix optimization problems, but it requires extensive knowledge and experience to formulate the model and it takes much time to solve it [6], [17]. A Linear programming mathematical model consists of the linear objective function under some constraints to guarantee the limitations of the system and non-negative decision variables [18]. However, better results might be obtained by customizing a mathematical model for a specific company [7]. For any company to manufacture any type of product, resources such as raw materials are required, and these resources are limited therefore they are production constraints [19]. For the soft variety product-mix problem, raw materials constraints exist to make sure that the consumed amount does not exceed the amount supplied. The resources that could be fully consumed and cause a delay in the production schedule are called bottle nicks or capacity constraint resources (CCRs) [20]. Thus, it is critical to define a performance measure for the CCRs. Most of the previous research solving product-mix problems under the Theory of Constraints (TOC) considered only one criterion [21] which is maximizing the profit. A producer should determine the optimum product mix that maximizes the profit, productivity, and raw materials' utilization [22] which are all taken into consideration in this study. The convenient profit calculation is by subtracting the expenses from the total revenue. It is important to consider measuring the performance over specific periods [23], so the performance measures are calculated on monthly basis in this study. Some previous studies compared LP a TOC in solving product mix problems and were found that both techniques have the same results as found in [24] and [25]. Four factors determine the product-mix decision, namely, the amount available of each resource, resources' costs, products' selling prices, and total market demand of each product as shown in Lea's linear programming model for conventional product-mix problem [26]:

$$\text{Maximize} \qquad Z = \sum_{j=1}^{n} (P_j - c_j) x_j \qquad (1)$$

$$\text{Subject to:} \quad \sum_{j=1}^{n} a_{ij} x_{ij} \leq b_i \qquad (2)$$

$$i = 1, 2, \dots, m \text{ (resource/capacity constraint)}$$

$$x_j \geq d_j \qquad (3)$$

$$j = 1, 2, \dots, n \text{ (resource/capacity constraint)}$$

$$x_i \geq 0 \qquad (4)$$

where $Z$ represents the objective function of maximizing profit, $P_j$ is representing product $(j)$ selling price, $c_j$ is the cost of production for the product $(j)$, $x_j$ is the number of the product $(j)$ to be produced, $a_{ij}$ represents the total amount of resource $(i)$ used for producing the product $(j)$, $b_i$ is representing the capacity of resource $(i)$, $d_j$ is the total demand of product $(j)$, the number of used resources is represented by $(n)$ and the number of products is represented by $(m)$.

Most of the previous product mix problems research work uses deterministic data while in real life, demand, production time, and capacity in many cases are uncertain that is the reason behind the need for manufacturing flexibility considering their values as intervals [6],[27]. One of the main flexibility types of manufacturing is the product-mix flexibility meaning producing a variety of products with minimum changeover costs [28], [29]. In short, a flexible optimization model is more beneficial when the resulting product-mix revenue is higher than the initial cost [30]. Although linear programming can solve many product-mix problems, many other product-mix problems have nonlinear objective functions such as having price flexibility that requires nonlinear programming [31]. Determining the optimum product mix in some cases may require multi-objective programming models or metaheuristics [32]. Other methodologies that are used for product mix problems are the activity-based costing method (ABC) developed by Kaplan and Cooper [11] and the time-driven activity-based costing (TDABC) [33].

## 2 PROBLEM STATEMENT

Company XYZ produces six products in a small manufacturing facility. One product is produced at a time and the setup time when changing from one product to another is negligible. All six products are competing on 27 types of raw materials that are allowed to be bought as multiples of specific amounts to benefit from the economies of scale. The basic raw materials amount and their prices are present in [41]. The product mix is soft, meaning that the same raw materials are used for producing the six products, but with different amounts for each product. The forecast of six months' demand and initial inventory for each product is shown in Table 1. The facility production capacity is 20800 packages/month. The selling prices of each type of product are present in Table 2. The costs related to production are the fixed cost of 88800 EGP/Month, the variable cost of 3.08 EGP/package produced, and the inventory cost of 3 EGP/package/month.

The objective is to develop a production schedule plan that satisfies the demand, maximizes the company's profit without incurring backlogs. This objective is by problem definition is coupled with determining the amounts of raw materials to be supplied for each month. Thus, it is concluded that two schedules should be the outcome of the problem, the 1st is for production and the 2nd is for materials supplies decision.

**Table 1: Demand required for each period of the six months**

| Periods ($t$) | A | B | C | D | E | F |
|---|---|---|---|---|---|---|
| **1.** | 4660 | 3256 | 3407 | 3966 | 5852 | 4531 |
| **2.** | 2982 | 3565 | 4914 | 2791 | 4031 | 5041 |
| **3.** | 3832 | 4574 | 3083 | 2873 | 3043 | 4748 |
| **4.** | 1293 | 3286 | 3993 | 2251 | 2990 | 4985 |
| **5.** | 1896 | 4748 | 3706 | 3550 | 2519 | 5167 |
| **6.** | 2357 | 3593 | 3327 | 3019 | 4125 | 3580 |
| **Initial Inv. ($I_0$)** | 3308 | 1839 | 2478 | 1673 | 3716 | 2164 |

**Table 2: Selling prices of a package of each product**

| Product Number ($i$) | Products Information | |
|---|---|---|
| | **Product Name** | **Price (EGP) / Package ($P_i$)** |
| 1. | A | 20 |
| 2. | B | 25 |
| 3. | C | 27 |
| 4. | D | 20 |
| 5. | E | 30 |
| 6. | F | 21 |

## 3    PROPOSED METHODOLOGIES

It is obvious from the problem definition that the problem is a multi-objective problem. The mathematical models presented in this section are developed to satisfy both objectives heuristically. This is in addition to satisfying all the constraints. At first, an integer linear programming heuristic is proposed giving a near-optimum solution.

The linear model developed here relaxes the raw materials doubling constraints allowing utilizing any fractional amount of raw material. After the model is optimized, the amounts of the materials are recalculated satisfying the doubling rule with assuming that whatever amount is left from a certain period can be used in the next period. This heuristic can give good solutions however, if the amounts of the raw materials are considered in the model as integer decision variables, it would be more rational and give the optimum solution. The model in this case would be very hard to solve in real-time. Thus, a nonlinear model is developed to overcome this issue and still give better solutions.

### 3.1    Proposed heuristic linear model

Any mathematical optimization model in operations research has three features to come with a decision. At first, the decision variables need to be specified. Then the problem constraints are defined restricting the feasible values of the decision variables. Finally, the objective function is constructed in terms of the decision variables to maximize or minimize the output [39]. Then business agents take their decision based on the outcomes of this function [40].

For the current mathematical model consider the following:

### 3.1.1 Heuristic inputs and notation

$i$: Products notation, $i = \{1, 2, \ldots, n = 6\}$

$j$: Raw materials, $j = \{1, 2, \ldots, m = 27\}$

$t$: Production periods, $t = \{1, 2, 3, \ldots, q = 6\}$

$S_i$: is the selling price for one package of the product $i$

$W_j$: is the weight in Kilograms of the initially determined material $j$

$W_{ji}$: is the weight required of raw material $j$ to produce one package of the product $i$

$W_{jt}$: is the weight required of raw material $j$ to meet the production schedule requirements at period $t$

$Q_t$: is the production capacity of the facility in period $t$

$D_{it}$: is the demand of product $i$ in period $t$

$h_t$: is the inventory cost per unit in period $t$

$C_j$: is the cost per Kilogram of the material (j)

$C_v$: is the variable production costs/package

$C_f$: is the fixed production costs

$E_{jt}$: is the actual amount of raw material $j$ at period $t$

$\lceil E_{jt} \rceil$: is the integer multiple of raw material $j$ required at period $t$

$B_{jt}$: is the difference between the amount to be bought of raw material $j$ and the actual amount required of raw material $j$ at period $t$

$UE_{jt+1}$: is the updated amount of raw materials needed for the period $(t + 1)$

$UZ$: is the updated profit function

### 3.1.2 Decision Variables

$X_{it}$ is the number of packages to be produced of product $i$ in period $t$

$I_{it}$ is the inventory level of product $i$ at period $t$

### 3.1.3 Constraints

1) Non-negativity Constraint:

$$I_{it}, X_{it} \geq 0 \quad \forall \, (i, t) \qquad (5)$$

2) Integrality Constraint:

$$I_{it}, X_{it} = int \qquad \forall \, (i, t) \qquad (6)$$

3) Production constraint

$$\sum_{i=1}^{n} X_{it} \leq Q_t \qquad \forall \, (t) \qquad (7)$$

4) Demand Constraint

$$I_{it-1} + X_{it} = D_{it} \quad \forall \, (i, t) \qquad (8)$$

### 3.1.4 Objective Function

Max Z= Revenue - Materials Cost –Inventory Costs – Variable Costs – Fixed Costs

$$\text{Max Z} = \sum_{i=1}^{n} S_i X_{it} - \sum_{j=1}^{m} C_j W_{jt} - \sum_{t=1}^{q} \sum_{i=1}^{n} h_t I_{it} - \sum_{i=1}^{n} C_v X_{it} - \sum C_F \quad (9)$$

$$\text{Where,} \, I_{it} = \sum_{t=1}^{t} X_{it} - \sum_{t=1}^{t} D_{it} + \sum_{t=0}^{t-1} I_{it} \quad \forall \, (I, t) \quad (10)$$

Steps of the proposed heuristic:

1. Solve the previous Mathematical model
2. Calculate the actual amounts needed (fraction)

$$E_{jt} = \frac{\sum_{i=1}^{6} W_{ji} X_{it}}{W_j} \qquad \forall \, (j, t) \qquad (11)$$

3. Round up for t=1 because a multiple of $W_j$ must be provided for the 1st period.

$$\lceil E_{j1} \rceil = \left\lceil \frac{\sum_{i=1}^{6} W_{ji} X_{i1}}{W_j} \right\rceil \quad \forall \, (j) \qquad (12)$$

4. Calculate the difference between the actual amount needed and the amount to be bought (Integer)

$$B_{jt} = \lceil E_{jt} \rceil - E_{jt} \qquad (13)$$

5. Update the amount needed for the next period

$$UE_{jt+1} = E_{jt+1} - B_{jt} \qquad (14)$$

6. Round up the updated amount

$$\lceil UE_{jt+1} \rceil = \lceil E_{jt+1} - B_{jt} \rceil \qquad (15)$$

7. Repeat the steps from 3 to 5 for t= 1,2, …, q-1
8. Calculate the actual cost of raw materials

$$\sum_{t=1}^{q-1} \sum_{j=1}^{m} C_j (\lceil E_{j1} \rceil + \lceil UE_{jt+1} \rceil) \qquad (16)$$

9. Recalculate the profit function by adding the difference between the actual cost of raw materials and the model cost of raw materials.

$$UZ = Z + \left( \sum_{j=1}^{m} C_j W_{jt} \right) - \sum_{t=1}^{q-1} \sum_{j=1}^{m} C_j W_j (\lceil E_{j1} \rceil + \lceil UE_{jt+1} \rceil) \qquad (17)$$

### 3.2 Proposed Non-Linear Mathematical Models

If we try to maximize the profit function (17) stated previously. The model becomes nonlinear as the function would have non-linear decision variables $(\lceil E_{j1} \rceil \, and \, \lceil UE_{jt+1} \rceil)$ that is non-linear. The objective function becomes:

$$\text{Max Z} = \sum_{i=1}^{n} S_i X_{it} - \sum_{t=1}^{q-1} \sum_{j=1}^{m} C_j (\lceil E_{j1} \rceil + \lceil UE_{jt+1} \rceil) - \sum_{t=1}^{q} \sum_{i=1}^{n} h_t I_{it} - \sum_{i=1}^{n} C_v X_{it} - \sum C_F \qquad (18)$$

A proven to be an efficient technique for solving Non-linear Programming problems is the Generalized Reduced Gradient method (GRG) [42]. It is an extension of Wolfe's reduced gradient method originally proposed to deal with nonlinear models with nonlinear constraints [43]. The mathematical model is solved using the Generalized Reduced Gradient (GRG) method integrated Microsoft Excel software with solver with multi-start option. This option allows the model to start solving using different starting points to find a probable optimum solution [44].

Since the GRG method doesn't guarantee the optimum solution, the authors thought that relaxing integer constraints could result in a better solution. In addition, this could reduce the computational power required for solving the problem.

## 4    RESULTS

In this section, the results of the proposed models are presented. First, the linear model is discussed in terms of the master schedule, profit, and raw materials utilization. Later, the non-linear model is discussed and a comparison between the two models is executed.

Computational experiments are executed on Microsoft Excel Solver software on a PC with Intel(R) Core(TM) i5-2450M CPU 2.50 GHz and 12.0 GB memory. The computational time for the linear model didn't exceed 60 seconds while for the non-linear model time varies between a few minutes to several hours depending on the starting solution and type of algorithm utilized.

### 4.1 Linear Heuristic Model Results

The model is initially modelled and solved via Microsoft Excel software and the objective function is optimized with total profit = 899044 for the production master schedule and inventory levels in Table 3 and Table 4, respectively. But after continuing the heuristic the updated profit decreased by 5% to be 853863 EGP. The utilization of the resources is found to be 96.07%.

### 4.2 Non-linear Model Results

Solving the non-linear model on excel solver software using the GRG method results are discussed in this section. GRG model has two options one start and multi-start. One start might cause the algorithm to be stuck in a local optimum. On the other hand, a multi-start option would guarantee at least a near-optimum solution but would require much computational cost.

When the model is solved using one start option it resulted in the production master schedule and inventory levels in Table. III and Table. IV, respectively. It took 40 minutes to achieve such a solution. The maximum profit obtained in this case is 923031 EGP and raw materials utilization obtained is 96.27%.

Relaxing the integer constraints over $X_{it}$ and $I_{it}$ results are present in Table 3 and Table 4, respectively. The solution improved and profit increased to become 929374 EGP while the utilization almost remained the same 96.36%. Furthermore, the solution time is reduced to be less than 2 minutes.

### 4.3 Discussion

The non-linear model solution takes an uncertain amount of time to obtain a good solution. However, the encounter increase in profit is ≈ 9%. Hence, the increase in profit worth solving the non-linear model.

It is noticed that for both models, the total amounts to be produced and total inventory levels for each period are the same. This can be noticed in Table 5. Each solution has a distinct production quantities and inventory for each period, resulting in a variance in the final profit value.

Table 3: Comparisons of the results of the three models master scheduling

| Period ($t$) | Models Master Scheduling Results | | | | | | | | | | | | | | | | | |
| | Integer Linear Model | | | | | | Integer Non-Linear Model | | | | | | Non-Linear Model (Integer Constraints Relaxed) | | | | | |
| | A | B | C | D | E | F | A | B | C | D | E | F | A | B | C | D | E | F |
| 1 | 1352 | 1417 | 2530 | 2293 | 2136 | 2367 | 2163 | 1418 | 929 | 2550 | 2138 | 2897 | 1352 | 1417 | 929 | 2293 | 2136 | 3968 |
| 2 | 2982 | 3565 | 1712 | 3468 | 4031 | 5041 | 2035 | 3563 | 4914 | 2277 | 4028 | 3982 | 2982 | 4242 | 4914 | 2791 | 4031 | 1839 |
| 3 | 3832 | 4574 | 3083 | 1519 | 3043 | 4748 | 2482 | 4574 | 3083 | 2873 | 3041 | 4746 | 3832 | 3220 | 3083 | 2873 | 3043 | 4748 |
| 4 | 1293 | 3286 | 3993 | 2644 | 2990 | 4985 | 1686 | 3286 | 3993 | 2251 | 2990 | 4985 | 1293 | 3286 | 3993 | 2644 | 2990 | 4985 |
| 5 | 1896 | 4748 | 3706 | 2764 | 2519 | 5167 | 1110 | 4748 | 3706 | 3550 | 2519 | 5167 | 1896 | 4748 | 3706 | 2764 | 2519 | 5167 |
| 6 | 2357 | 3593 | 3327 | 3019 | 4125 | 3580 | 2357 | 3593 | 3327 | 3019 | 4125 | 3580 | 2357 | 3593 | 3327 | 3019 | 4125 | 3580 |
| Profit | 853,863 | | | | | | 923,031 | | | | | | 929,374 | | | | | |
| $\sum c_j$ | 1,365,098 | | | | | | 1,306,571 | | | | | | 1,295,808 | | | | | |
| Utilization | 96.07% | | | | | | 96.27% | | | | | | 96.36% | | | | | |

**Table 4: Comparisons of the results of the three models inventory levels**

| Period (*t*) | Models Inventory Levels Results | | | | | | | | | | | | | | | | | |
|---|---|---|---|---|---|---|---|---|---|---|---|---|---|---|---|---|---|---|
| | Linear Model | | | | | | Non-Linear Model | | | | | | Non-Linear Model (Integer Constraints Relaxed) | | | | | |
| | A | B | C | D | E | F | A | B | C | D | E | F | A | B | C | D | E | F |
| 1 | 3308 | 1839 | 2478 | 1673 | 3716 | 2164 | 3308 | 1839 | 2478 | 1673 | 3716 | 2164 | 3308 | 1839 | 2478 | 1673 | 3716 | 2164 |
| 2 | 0 | 0 | 1601 | 0 | 0 | 0 | 0 | 0 | 0 | 0 | 0 | 1601 | 0 | 0 | 0 | 0 | 0 | 1601 |
| 3 | 0 | 0 | 0 | 677 | 0 | 0 | 0 | 677 | 0 | 0 | 0 | 0 | 0 | 677 | 0 | 0 | 0 | 0 |
| 4 | 0 | 0 | 0 | 0 | 0 | 0 | 0 | 0 | 0 | 0 | 0 | 0 | 0 | 0 | 0 | 0 | 0 | 0 |
| 5 | 0 | 0 | 0 | 393 | 0 | 0 | 0 | 0 | 0 | 393 | 0 | 0 | 0 | 0 | 0 | 393 | 0 | 0 |
| 6 | 0 | 0 | 0 | 0 | 0 | 0 | 0 | 0 | 0 | 0 | 0 | 0 | 0 | 0 | 0 | 0 | 0 | 0 |
| Total Cost | 53547 | | | | | | 53547 | | | | | | 53547 | | | | | |

**Table 5: Total production and inventory within each period**

| Period | Total Production | Total Inventory |
|--------|-----------------|-----------------|
| 1. | 12095 | 15178 |
| 2. | 20799 | 1601 |
| 3. | 20799 | 677 |
| 4. | 19191 | 0 |
| 5. | 20800 | 393 |
| 6. | 20001 | 0 |

From Table 4. it can be noticed that the inventory costs of the linear model and the non-linear models are the same. Furthermore, since the total production of each period is the same, then the variable cost will be the same also. Again, by definition of the problem, the revenue and the fixed costs are the same for all 3 models. This leaves us with one term that is affecting the profit and caused its raw materials costs.

It is expected that the linear model would be the least efficient in the cost of the raw materials estimation matter. This is because it doesn't consider the raw materials while optimizing the model, but rather the materials costs are updated after the model is solved. On the other hand, the non-linear models consider the amounts of the raw materials as decision variables. The difference in raw materials costs can be noticed in Table 3. This relaxation of the integer constraints in the 3$^{rd}$ model made it easier to find better results in terms of raw materials amounts that should be supplied. Thus, it gave the best results among the three models.

In both models the inventory of any month was consume directly in the next month, and that make sense where the inventory penalty increases with time.

Better solutions were obtained when the production and inventory sequences for all months from the linear model were applied to the nonlinear model, and even better solutions were obtained when the integer constraint was relaxed at a rate of roughly 9% higher than the linear model solution.

## 5    CONCLUSION

In this work, an integer (linear and non-linear optimization models), as well as a basic non-linear model are developed. This is to solve the product mix under study and come out with a proper production master schedule. In addition, the problem aims to determine the amounts of raw materials to be supplied for each period. Furthermore, the inventory levels are another thing that needed attention. Thus, the problem is considered as a multi-objective problem.

Finally, the results of the three models were compared to each other in terms of profit, raw materials costs, inventory costs, inventory levels, and raw materials utilization. In the three cases, obtained solutions satisfied the problem's requirements and constraints. However, the quality of the solution improved by 9% in terms of profit when the 3rd model is applied compared to the linear heuristics results. This is consistent with the decrease of raw materials cost. It is concluded that the 3$^{rd}$ model is the best model to use in cases like the one considered in this paper.

Future development of this study is to consider a higher number of products and check for the solution quality and time. Sensitivity analysis can be also conducted to check the robustness of the models utilize.

## 6 ACKNOWLEDGMENT

The authors wish to acknowledge the support of King Fahd University of Petroleum and Minerals.

The authors are grateful for Mr. Ahmed Fathy who provided the data of the problem that is processed in this paper. Special thanks to Dr. Sally Kassem, who co-authored a previous paper on the same problem with different topic. This work is an ongoing charity on behalf of our families, may Allah grant them peace.